\documentclass[twocolumn, showpacs,preprintnumbers,amsmath,amssymb]{revtex4}

\usepackage{graphicx}
\usepackage{dcolumn}
\usepackage{bm}

\begin{document}

\title{ Quantum Phase transition under pressure in a heavily hydrogen-doped iron-based superconductor LaFeAsO
  }

\author{  Naoki Fujiwara$^{1*}$, Naoto Kawaguchi$^{1}$, Soushi IImura$^{2}$, Satoru Matsuishi$^{3}$, Hideo Hosono$^{2, 3}$}

\affiliation{$^1$ Graduate School of Human and Environmental Studies,
Kyoto University, Yoshida-Nihonmatsu-cyo, Sakyo-ku, Kyoto 606-8501,
Japan}

\email { naoki@fujiwara.h.kyoto-u.ac.jp}

\affiliation {$^2$ Institute for Innovative Research, Tokyo
Institute of Technology, 4259 Nagatsuda,  Midori-ku, Yokohama
226-8503, Japan}

\affiliation{$^3$ Materials Research Center for Element Strategy, Tokyo Institute of
Technology, 4259 Nagatsuda, Midori-ku, Yokohama 226-8503, Japan }


\begin{abstract}
{ Hydrogen (H)-doped LaFeAsO is a prototypical iron-based superconductor. However, its phase diagram extends beyond the standard framework, where a superconducting (SC) phase follows an antiferromagnetic (AF) phase upon carrier doping; instead, the SC phase is sandwiched between two AF phases appearing in lightly and heavily H-doped regimes. We performed nuclear magnetic resonance (NMR) measurements under pressure, focusing on the second AF phase in the heavily H-doped regime. The second AF phase is strongly suppressed when a pressure of 3.0 GPa is applied, and apparently shifts to a highly H-doped regime, thereby a "bare" quantum critical point (QCP) emerges. A quantum critical regime emerges in a paramagnetic state near the QCP, however, the influence of the AF critical fluctuations to the SC phase is limited in the narrow doping regime near the QCP. The optimal SC condition ($T_c \sim48$ K) is unaffected by AF fluctuations.}
\end{abstract}

\pacs{74.25.Dw, 74.25.Ha, 74.20.-z, 74.25.nj}

\maketitle

Since the discovery of superconductivity in the iron-based pnictide LaFeAsO$_{1-x}$F$_x$ (F-doped La1111 series) [1], extensive studies on various types of iron-based compounds have been conducted. To date, the Sm1111 series marks the highest superconducting (SC)-transition temperature ($T_c$) ($T_c$=55 K) [2]. However, the 122 series has been mainly examined, as large single crystals are readily available [3, 4]. For the 122 series, an antiferromagnetic (AF) phase is followed by an SC phase with increasing carriers. The coexistence of these two phases near the phase boundary suggests an intimate interrelationship, indicating that AF fluctuations may be deeply associated with the SC mechanism.

\begin{figure*}
\includegraphics[scale=0.8]{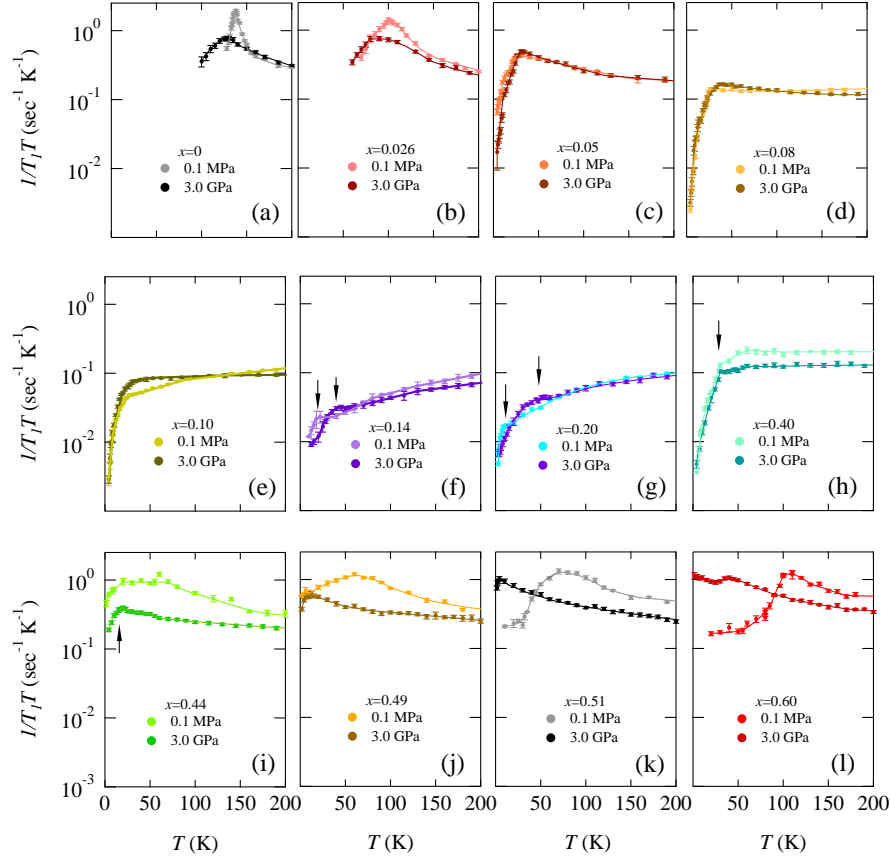}
\caption{\label{fig:epsart} {Evolution of  $1/T_1T$ for $^{75}$As with respect to F or H doping levels (denoted by $x$) at ambient pressure (0.1 MPa) and 3.0 GPa.} The sharp peaks in Figs. 1(a) and 1(b) indicate the Neel temperature, $T_N$. The arrows in Figs. 1(f), 1(g), 1(h), and 1(i) indicate the SC transition temperature. In a heavily H-doped regime, the second AF phase emerges accompanied with a round peak of  $T_N$ (Figs. 1(j), 1(k), and 1(l)).  }
\end{figure*}

However, exotic phase diagrams have been discovered for several compounds, such as the H-doped R1111 series (R=La, Ce, Sm, Gd) [5, 6], the P-doped La1111 series [7], FeSe [8-10], and R$_{1-x}$Fe$_{2-y}$Se$_2$ (R=K and Tl$_{0.6}$Rb$_{0.4}$) [11]. The phase diagrams of these compounds are not explainable within a conventional framework, where a single SC phase follows a single AF phase by tuning parameters such as the doping level or pressure. Among these examples, the H-doped R1111 series is relatively simple because chemical tuning is performed only via the hydrogen doping of RO layers; consequently, the FeAs layers are free from local disturbance and randomness. Hydrogen can be doped in a wide doping range, leading to a new AF phase in a heavily doped regime [12-14]. That is, the SC phase is sandwiched between two AF phases in the phase diagram. Note that the H-doped R1111 series (R= Ce, Sm, Gd) has an SC dome with a $T_c$ ranging from 45 to 50 K [6]. The SC dome moves to a lightly H-doped regime in the order of Ce, Sm, and Gd.  Unlike these compounds, the La1111 series exhibits double domes with a minimum $T_c$ of 15 K. Upon applying pressure, the double domes transform into a single dome [5, 15] with a high $T_c$ like the other R1111 series (R= Ce, Sm, Gd). This single dome moves to a lightly H-doped regime under increasing pressure, suggesting that the application of pressure is equivalent to the R replacement with heavier elements. For instance, the SC dome observed for the La1111 series at 3.0 GPa is almost identical to that for the Ce1111 series at ambient pressure (0.1 MPa). The application of pressure is more advantageous than the R replacement, as the influence of magnetic R elements is excluded.

We investigated an interrelationship between the SC phase and the second AF phase in a heavily H-doped regime via nuclear magnetic resonance (NMR) measurements for $^{75}$As, by comparing the phase diagram at 3.0 GPa with that at 0.1 MPa.

$^{75}$As-NMR measurements for the powder samples were acquired using a conventional coherent-pulsed NMR spectral meter. The relaxation rate ($1/T_1$) was measured using a conventional saturation-recovery method. $^{75}$As-NMR spectra for the powder samples exhibit double edges in the field-swept spectra due to the quadrupole interaction, and  $1/T_1$ was measured at the lower-field edge, where the FeAs planes are parallel to the applied field. A pressure of 3.0 GPa was applied using a NiCrAl-hybrid clamp-type pressure cell. We used a mixture of Flouorinert FC-70 and FC-77 as the pressure-mediation liquid. A coil wounded around the powder samples and an optical fiber with the Ruby powders glued on top were inserted into the sample space of the pressure cell. The pressure was monitored through Ruby fluorescence measurements [16].

\begin{figure*}
\includegraphics[scale=0.8]{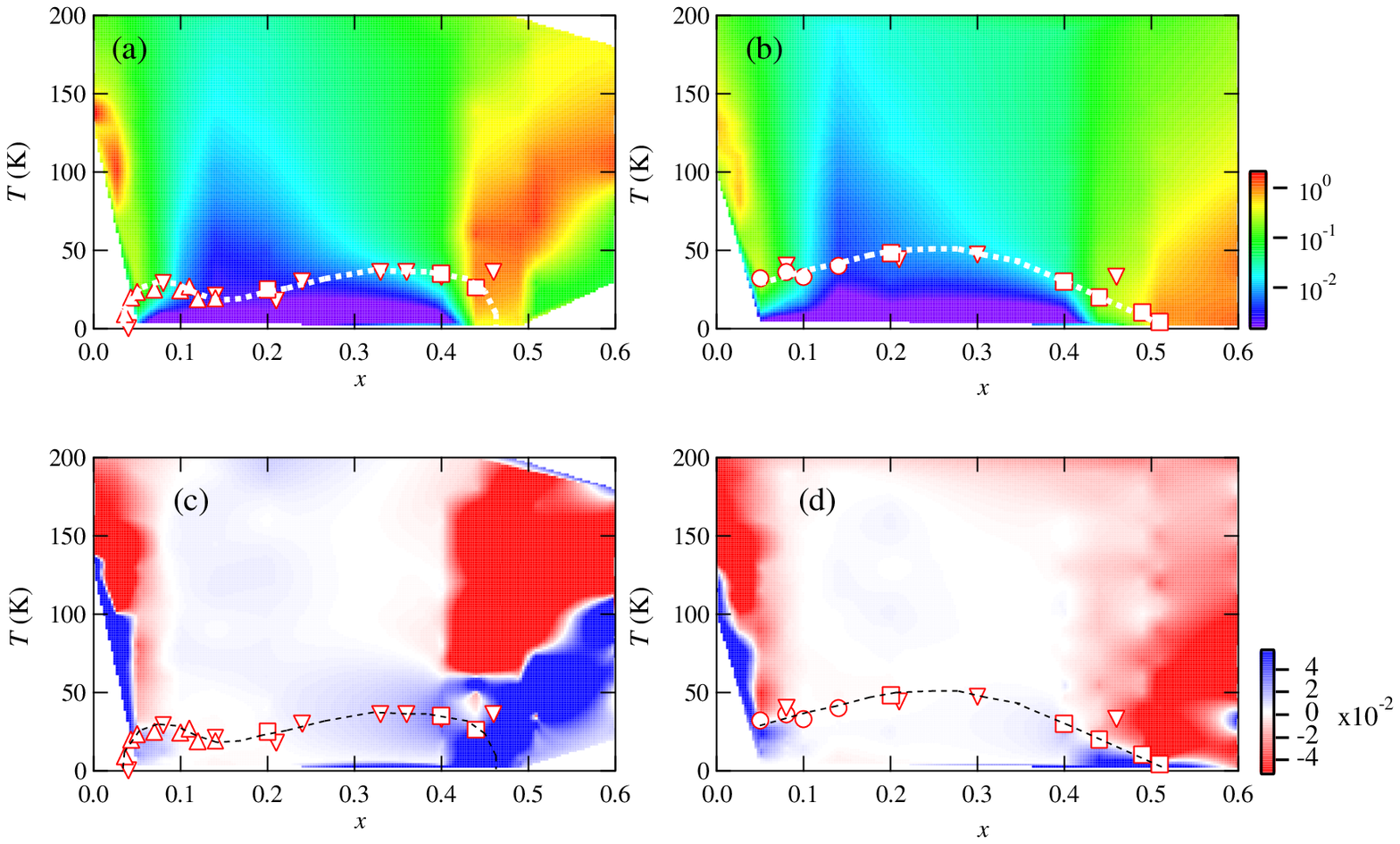}
\caption{\label{fig:epsart} Color maps of $1/T_1T$ and $\frac{d}{dT}(1/T_1T)$. Circles and squares represent data determined from $1/T_1T$ for F- and H-doped samples, respectively. Upright and inverted triangles represent data determined from the resistivity for  F- and H-doped samples, respectively [1, 6]. (a) and (b) $1/T_1T$ measured at 0.1 MPa and 3.0 GPa, respectively. (c) and (d) $\frac{d}{dT}(1/T_1T)$ measured at 0.1 MPa and 3.0 GPa, respectively.    }
\end{figure*}

\begin{figure}
\includegraphics[scale=0.6]{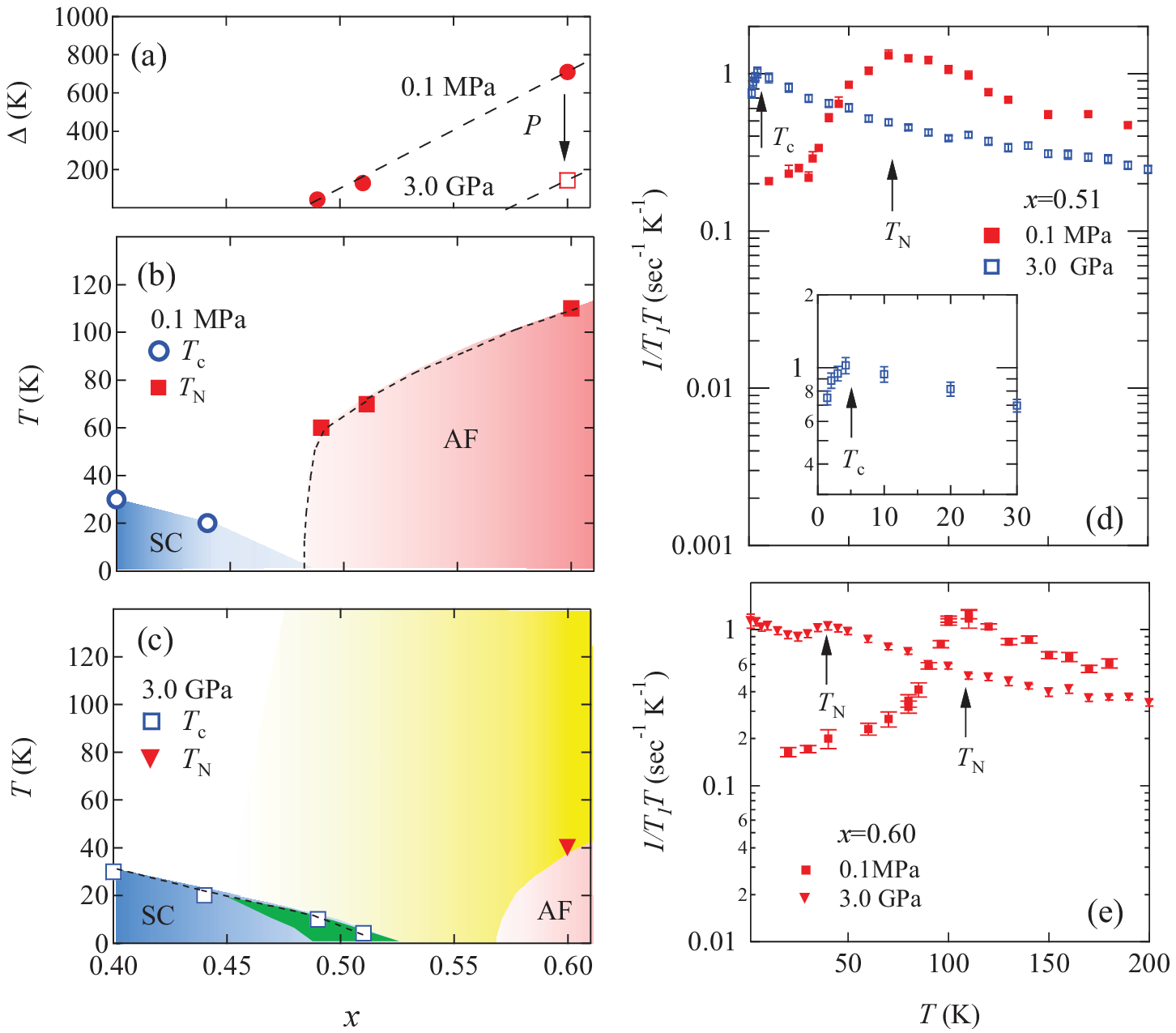}
\caption{\label{fig:wide} Excitation gap, phase diagrams, and $1/T_1T$ for a heavily H-doped regime. (a) The excitation gap $\Delta$ is estimated from Eq. (3). (b) Electronic phase diagram for a heavily H-doped regime determined from the $1/T_1T$ of $^{75}$As at 0.1 MPa. (c) The phase diagram determined from the $1/T_1T$ of $^{75}$As at 3.0 GPa. The portion colored in green shows the SC phase induced under pressure. The paramagnetic state in yellow shows the quantum critical regime. (d) $1/T_1T$ of $^{75}$As for 51\% H-doped samples. The inset is an expansion of $1/T_1T$ measured at 3.0 GPa. The arrows indicate the superconducting transition temperature ($T_c$) and the Neel temperature ($T_N$). (e) $1/T_1T$ of $^{75}$As for 60\% H-doped samples.  }
\end{figure}

The relaxation rate divided by temperature ($T$), written as $1/T_1T$, provides a measure of low-energy magnetic fluctuations. According to a theoretical investigation regarding two-dimensional AF systems [17], $1/T_1T$ exhibits a Curie-Weiss upturn with decreasing temperature toward the Neel temperature ($T_N$), and diverges or adopts a maximum at $T_N$.

\begin {equation}
\frac{1}{T_1T} \sim \frac{1}{T-T_N} + const.
\end {equation}
 For the case where AF fluctuations are weak or absent, $1/T_1T$ no longer exhibits the upturn, but instead, obeys the Korringa relation, which is proportional to the square of the density of states (DOS):

\begin {equation}
\frac{1}{T_1T} \sim \sum_{i} n_i(\varepsilon_F)^2,
\end {equation} where $n_i(\varepsilon_F)$ represents the DOS for the $d_{xy}, d_{yz}$, and $d_{zx}$ orbits of Fe ions.

Figures 1(a)-1(l) show the evolution of $1/T_1T$ measured at both 0.1 MPa and 3.0 GPa for several F or H doping levels.  For F doping, the maximum doping level is less than $x$=0.14-0.15 [18-20]. First, we focus on $1/T_1T$ at 0.1 MPa. For undoped and poorly F-doped regimes ($x <$ 0.05), $1/T_1T$ exhibits an upturn toward $T_N$ and a sharp peak at $T_N$, reflecting Eq. (1).  As the doping level increases ($x$=0.05-0.10), the AF phase vanishes, however, the AF fluctuaions remain. The upturn corresponding to  AF fluctuations is monotonically suppressed with increasing doping level. At doping levels of $x$=0.14-0.20, $1/T_1T$ no longer exhibits the upturn, but instead exhibits a steady decrease with decreasing temperature, except for a plateau just above $T_c$ (Figs. 1(f) and 1(g)) [18, 21]. This steady decrease originates from the loss of the DOS (see Eq. (2)). The loss of the DOS has been observed in photon-emission spectroscopy measurements [22-24], and is known as pseudo-gap behavior.  Upon further H doping ($x$=0.4-0.44), AF fluctuations return above $T_c$. In the heavily H-doped regime ($x\geq$ 0.49), the second AF phase emerges with an accompanying round peak at $T_N$ (Figs. 1(j)-1(l)).

When pressure is applied, remarkable features appear in two regimes: $x$=0.14-0.20 and $x\geq$ 0.49. For $x$=0.14-0.20, the minimum $T_c$ at 0.1 MPa changes the maximum $T_c$ and reaches 48 K, which is comparable to the highest $T_c$ of 55 K (see Fig. 1(g)).  For $x\geq$ 0.49, the second AF phase is strongly suppressed when pressure is applied. These remarkable features are clearly seen from the color maps in Figs. 2(a)-2(d).

The absolute values of $1/T_1T$ at 0.1 MPa and 3.0 GPa are plotted in Figs. 2(a) and 2(b), respectively, and the derivatives of $1/T_1T$ with respect to temperature, i.e., $\frac{d}{dT}(1/T_1T)$, at 0.1 MPa and 3.0 GPa are plotted in Figs. 2(c) and 2(d), respectively. In a paramagnetic state above $T_N$, $\frac{d}{dT}(1/T_1T)$ becomes negative (shown in red), whereas for both the SC and AF states, $\frac{d}{dT}(1/T_1T)$  becomes positive due to the decrease of quasi-particles or magnetic excitations at low temperatures. The SC domes appear to have no direct correlation with AF fluctuations or AF phases. Both the minimum $T_c$ at 0.1 MPa and the maximum $T_c$ at 3.0 GPa are observed within the same region, where the absolute values of $1/T_1T$  are small and AF fluctuations are absent, shown in blue in Figs. 2(a) and 2(b). The areas in blue also correspond to the region where the resistivity of the powder samples exhibits a $T^2$ temperature dependence [25], suggesting a Fermi liquid state.

Detailed phase diagrams for the heavily H-doped regime ($x\geq$ 0.49) are shown in Figs. 3(b) and 3(c). The excitation gap  $\Delta$  of the AF state (Fig. 3(a)) is estimated using the following formula:
\begin {equation}
 \frac{1}{T_1T} \sim e^{-\Delta/T}+const .
\end {equation} The gap $\Delta$ represents the order parameter of the AF phase, and becomes zero at $x\sim$0.49 at 0.1 MPa. The data of $T_c$ and $T_N$ in Figs. 3(b) and 3(c) are determined based on the $T$ dependence of $1/T_1T$. In Fig. 3(d), the round peak corresponds to $T_N$, whereas the onset of the sharp drop below 4 K corresponds to $T_c$. The second AF phase is strongly suppressed at 3.0 GPa, and apparently shifts to a highly H-doped regime in the $x-T$ phase diagram. For $x$=0.49 and $x$=0.51, a quantum phase transition form the AF to SC phases occurs; the AF phase with a $T_N$ of 60-70 K vanishes, while instead the SC phase accompanied by a very low $T_c$ emerges at 3.0 GPa.  The phase transition from the AF to SC phases no longer occurs for $x \geq 0.6$ (see the round peaks in Fig. 3(e)).

 The phase diagram at 3.0 GPa, as illustrated in Fig. 3(c), is highly important with respect to the quantum critical point (QCP) where the AF state disappears. The AF state disappears at the QCP due to a mismatching of the nesting conditions between electron pockets [26, 27] and/or broadening of the d-orbital bandwidth [28]. A "bare" QCP emerges under pressure owing to the phase segregation between the AF and SC phases. A fan-shaped nobel quantum critical regime is theoretically predicted for a wide $T$ region near the QCP [28, 29], as highlighted in yellow in Fig. 3(c). The critical regime appears as the region where $1/T_1T$ exhibits Curie-Weiss-like behavior, as shown in Figs. 1(i), 1(j), 1(k), and 3(d): $1/T_1T \propto \frac{1}{T-\theta} + const$ where $\theta$ is -52 K for both 49\% and 51\% doped samples. Furthermore, the critical fluctuations mix with the gapped excitations of the AF state near the QCP, which can be observed as the upturn of $1/T_1T$ below $T_N$ in Figs. 1(l) and 3(e). The upturn contrasts to the downturn due to opening of the superconducting gap, as shown in the inset of Fig. 3(d).

The pressure-induced SC phase emerges in the narrow doping regime where the AF phase vanishes under pressure (see the area in green in Fig. 3(c)). The pressure-induced SC phase is accompanied by the critical regime above $T_c$, implying that the AF critical fluctuations would be a key factor for the pairing interaction. However, the SC phase away from the QCP is almost unchanged ($x$=0.40 and 0.44), implying that the influence of the AF critical fluctuations is limited in the narrow doping regime where the superconductivity is induced under pressure. This result is consistent with the fact that this system achieves the highest $T_c$ without the influence of AF fluctuations for $x$=0.14-0.20 [18, 21]. The experimental result is advantageous for the orbital-fluctuation-mediated mechanism [30, 31], which is a counterpart of the AF-fluctuation-mediated mechanism. These results imply that there are competing pairing interactions within the SC phase. If AF fluctuations were solely involved in the $T_c$ optimal condition, the optimal doping level at 3.0 GPa would shift to a higher doping regime near the QCP following the apparent shift of the second AF phase, and the SC double-domes structure observed at 0.1 MPa would become much more prominent at 3.0 GPa. To determine whether the superconducting mechanism is of magnetic origin or of orbital origin in the whole doping regime, an interrelationship between the SC and AF phases is important. The SC dome and the second AF phase move to the opposite directions in the $x-T$ phase diagram under increasing pressure, leading to the expansion of the quantum critical regime. Therefore, the behavior of the SC phase colored in green in Fig. 3(c) under higher pressure than 3.0 GPa would provide an important clue to determine the pairing symmetry.

In summary, we have studied the second AF phase of H-doped LaFeAsO under pressure via $^{75}$As-NMR techniques. The second AF phase is strongly suppressed under pressure, thereby a "bare" QCP emerges.  A pressure-induced quantum phase transition from the AF to SC phases occurs in a narrow doping regime near the QCP. The quantum critical regime emerges in a wide $T$ region above the QCP. However, the influence of the AF critical fluctuations to the SC phase is limited in the narrow doping regime near the QCP. The optimal $T_c$ condition is unaffected by AF fluctuations, implying the existence of competing pairing interactions in the SC phase.

H. Hosono acknowledges a fund from the MEXT Element Strategy Initiative to form a Research Core.

\ \\

\end{document}